\documentclass[aps,preprint,amsmath,amssymb,superscriptaddress,floatfix]{revtex4-1}
\usepackage{mathrsfs}
\usepackage{array}
\usepackage{amsfonts}
\usepackage{amsmath}
\usepackage{amstext}
\usepackage{amssymb}
\usepackage{bbold}
\usepackage{bm}
\usepackage{color}
\usepackage{graphicx}
\usepackage{epstopdf}
\usepackage{mathrsfs}
\begin{document}
%
\title{Exact Master Equation and General Non-Markovian Dynamics\\ in Open Quantum Systems}
\author{Wei-Min Zhang\footnote{Invited mini-review based on the talk presented in the Conference: \emph{Frontiers of 
Quantum and Mesoscopic Thermodynamics}, Prague, Czech Republic, July 9-15 (2017); Email: wzhang@mail.ncku.edu.tw} }
\affiliation{Department of Physics and Center for Quantum Information Science, National Cheng Kung
University, Tainan 70101, Taiwan}
\begin{abstract}
Investigations of quantum and mesoscopic thermodynamics force one to answer two 
fundamental questions associated with the foundations of statistical mechanics:
(i) how does macroscopic irreversibility emerge from microscopic reversibility? (ii) how does
the system relax in general to thermal equilibrium with its environment? The answers to these
questions rely on a deep understanding of nonequilibrium dynamics of systems interacting 
with their environments. Decoherence is also a main concern in developing quantum information
technology. In the past two decades, many theoretical and experimental investigations 
have devoted to this topic, most of these investigations take the Markov (memory-less) approximation. 
These investigations have provided a partial understanding to several fundamental issues, 
such as quantum measurement and the quantum-to-classical transition, etc.
However, experimental implementations of nanoscale solid-state quantum information processing 
makes strong non-Markovian memory effects unavoidable, thus rendering their study a pressing 
and vital issue.  Through the rigorous derivation of the exact master equation and a systematical
exploration of various non-Markovian processes for a large class of open quantum systems, we find 
that decoherence manifests unexpected complexities.  We demonstrate these general non-Markovian 
dynamics manifested in different open quantum systems.
\end{abstract} 
\maketitle
\section{Introduction}
\label{intro}
As it is well-known, any realistic system will inevitably interact with its environment. 
For nano-scale quantum devices or more general mesoscopic systems, such interactions are usually not 
negligible, and thus these objects must be treated as open systems. Specifically, an open system is 
defined as the principal system consisting of only a few relevant dynamical variables in contact 
with an  environment containing a huge (infinite) number of degrees of freedom.
Understanding the dynamics of open systems is also one of the most challenging 
topics in physics, chemistry, biology and even engineering.  In particular, the interactions between the 
principal system and the environment can induce various back-reactions between them such that the 
system can memory its historical evolution. These memory processes are characterized as the 
non-Markovian dynamics in open systems, to distinguish from the memoryless processes 
which are named as Markov dynamics in the literature. 

Physically, non-Markovian dynamics can be described by different time correlations 
associated with environment-induced dissipation and fluctuation dynamics  in open 
 systems. To understand the quantum dynamics of open  systems, many different 
approaches were developed.  In principle,  the system we care about plus its environments 
together form a closed system, which is governed by the Schr\"{o}dinger equation in terms 
of the wave function of the total system or the von Neumann equation in terms of 
the total density matrix.  Usually the environment is assumed to be initially in a 
thermal equilibrium state at a given temperature $T$, which is a mixed state. Thus, the 
Schr\"{o}dinger picture is no longer applicable.  One has to use the von Neumann 
equation in terms of the density matrix to solve the dynamics of the total system. 
The solution of the density matrix contains all predictions to various physical observables.  

Practically, it is very difficult to solve the von Neumann equation of the
total system, due to the infinite number of degrees of freedom involved in the environment. 
More important, we are only interested in the dynamics of the principal system itself, rather than 
the dynamics of its environment.  Hence, for a long time, a central issue in the investigations of
the dynamics of open system has been focused on finding the equation of motion for 
the reduced density matrix of the principal system. Such an equation of motion is  called 
as {\it Master Equation}. Within the framework of von Neumann equation of the
total system, the master equation can be derived in principle from quantum mechanics, as I will attempt 
to do so in this article. On the other hand, from the open system point of view, the master equation 
actually plays an even more important role for the foundation of statistical mechanics \cite{Huang87},  
in comparison with the Newtonian equation for macroscopic objects,  the Maxwell equations for 
electrodynamics, and the Schr\"{o}dinger equation for isolated quantum systems. From the more 
fundamental point of view, the Newtonian equation can be derived from the Lagrangian formalism, 
the Maxwell equations can be derived from the field theory of quantum electrodynamics (QED), and the
Schr\"{o}dinger equation is only a nonrelativistic approximation of the Dirac equation 
which can also be derived from QED, while it is not clear if there is a fundamental principle 
to directly determine the master equation. With this aspect, finding the master equation for open quantum 
systems is a big challenge in science. Certainly, if one can find the exact master equation
for arbitrary open systems, many interesting and fundamental problems in open system dynamics, 
including non-Markovian memory dynamics I will focus on this in article, can be easily addressed.

Historically, the first master equation was phenomenologically introduced by Pauli in 1928 \cite{Pauli}, 
which is now called the Pauli master equation in the literature.  In the past many decades,
one has made progresses with various approaches in deriving the master equation 
for  different open quantum systems. These include the Nakajawa-Zeneng master equation 
\cite{Nakajima,Zwanzig}, the Born or Born-Markov master equation \cite{Hove,Haake}, 
the GKS-Lindblad master equation \cite{GKS,Lindblad}, etc. However, all these master equations 
are either practically unsolvable \cite{Nakajima,Zwanzig} or only applicable for Markov dynamics 
\cite{Hove,Haake,GKS,Lindblad}.  Until 1980's, Caldeira and Leggett systematically derived a master 
equation for quantum Brownian motion \cite{Caldeira1983}, using the Feynman-Vernon influence 
functional approach \cite{Fey1963} to explicitly and exactly integrate out all the environment 
degrees of freedom. Since then, the quantum Brownian motion becomes a prototype example
in understanding the dynamics of open quantum systems \cite{Weiss08}. 
In reality, however, no many systems can be treated with the quantum Brownian motion. There is still a lack 
of satisfactory answer to the master equation and to open quantum system dynamics in general. 
Indeed, not having a rigorous and more general master equation for a large class of realistic open
quantum systems remains a primary obstacle in understanding many fundamental problems in physics.


To address physically the more general essence of open quantum system dynamics, 
we developed a full non-Markovian decoherence theory \cite{PRL2012,PRB2008,NJP2010,ANNP2012,PRB2015,PRB18} 
with the exact master equation we derived recently for a class of open systems. These open systems 
linearly couple to environments but are different from the quantum Brownian motion. More specifically, the 
system and the environments are described with Fano-Anderson type Hamiltonians \cite{Anderson1961,Fano1961} 
that have wide applications in atomic physics, quantum optics, condensed matter physics and 
particle physics \cite{Tannoudji92,Lam00,Mahan2000}.  Here both the system and the environment 
can be either bosonic or fermionic, and it may also be extendible to spin-like systems. 
We find \cite{PRL2012,PRB2008,NJP2010,ANNP2012,PRB2015,PRB18} that 
the dissipation and fluctuations coefficients in our exact master equation are intimately connected
with nonequilibrium Green functions in many-body systems \cite{Schwinger1961,Keldysh1965,Kadanoff1962}. 
As a result, we show that the nonequilibrium Green functions that obey the integro-differential convolution 
equations depict all possible non-Markovian memory dynamics through the time-convolution integral structures
\cite{PRL2012,Xiong2010,Sci2015b,Sci2015a,PRA2015,Lei2011,Ali2017}.  

I should emphasize from the very beginning that in the reality, non-Markovian 
dynamics is well-defined as memory processes in open systems. Experimentally, 
non-Markovian dynamics can be quantified though direct measurements of 
two-time correlation functions which demonstrate explicitly memory effects. 
The fundamental study of quantum dynamics and the technology development of 
quantum information processing show that it is crucially important to understand  general 
physical behaviors of non-Markovian dynamics, namely how different energy structures of the 
system and the environment, and different couplings between them, including different initial 
state dependences on the system and the environment, determine different memory effects 
of open quantum systems. Such understanding could truly help one to engineeringly control and 
manipulate decoherence in practical applications in quantum technology, and therefore 
it is my main concern in the study of non-Markovian dynamics in open quantum systems.

It may also be worth pointing out that the study of non-Markovian decoherence dynamics in open quantum 
systems has attracted a great deal of attentions recently \cite{Vega2017}. Most of investigations 
have been focused on how to mathematically quantify the degree of non-Markovianity by introducing 
different concepts, such as divisibility \cite{Wolf2008,Rivas2010} and distinguishability of states 
\cite{Breuer09}, etc., in an attempt of mathematically characterizing quantum Markovianity. The
results from these investigations are much definition-dependent and  the conclusions often diverge 
from each other. It is therefore not the topic I will discuss in this article. 

 \section{Structures of system-environment couplings}
 
Undoubtedly, dynamics of open quantum systems crucially depend on the structure of
system-environment couplings or interactions. Here I would like to begin with a discussion 
on possible system-environment couplings we may encounter in practical applications.

A general open quantum system is defined as a principal system of interest interacting 
with its surroundings as its environments (or reservoirs).
The interactions between the system and its environments are the manifestation of   
the exchanges of matters, energies and informations between them.  A general Hamiltonian 
describing the coupling between the system and its environments can be formally written as
\begin{align}
H=H_S + H_E + H_{\rm SE}, 
\end{align}
where $H_S$ and $H_E$ are the Hamiltonians of the system and the environment,
respectively, and $H_{SE}$ denotes the interaction between them.
The explicit form of $H_{SE}$ depends on the particular system and its environment(s) 
that we concern.   

The simplest realization of such an open quantum system was originally introduced 
by Feynman and Vernon in the seminal paper in 1963 \cite{Fey1963}, where Feynman and 
Vernon developed a theory named influence functional, based on the path-integral formalism 
\cite{Fey65}, to deal with the influence of linear environmental systems acting on the principal 
system. In particular, they modeled the linear environmental 
systems as a sum of an infinite number of harmonic oscillators.
In 1980s, Caldeira and Leggett used the approach of Feynman-Vernon influence functional to 
study in details the dynamics of a Brownian particle under the influence of 
such an environment  \cite{Caldeira1983},  which is now called as the Caldeira-Leggett (CL) model in the literature. 
The significance of the CL model is the emergence of the classical dissipation 
motion of a Brownian particle within the framework of quantum mechanics. 
The equilibrium fluctuation-dissipation theorem proposed originally by Einstein is also
naturally obtained quantum mechanically  from this model.  
Thus, the CL model becomes a prototype model in the study of the dynamics 
of open quantum systems \cite{Weiss08}.  

On the other hand, the rapid developments of  new emerging research fields on nano technologies and 
quantum information sciences have stimulated tremendous
interests on decoherence dynamics of tiny quantum devices, due to their invertible interaction with
various environments surrounding.  Typical examples include semiconductor nanostructures 
in mesoscopic physics, nanophotonics in photonic crystals and metamaterials, 
nanostructured cavity QED, 
and electrons or nuclear spins interacting with a thermal spin bath, just to name a few.  These open quantum 
systems, which are currently very popular in the practical applications of nanotechnologies and quantum information
sciences, go much beyond the Caldeira-Leggett model of the quantum Brownian motion. 
The general Hamiltonian for these system-environment couplings should be analyzed physically in a 
more realistic manner, as I will discuss in the following. 

Consider first the exchange of matters between the system and its environments. Since matters are built 
with fermions, i.e. electrons and nuclei (or more fundamentally, quarks), we should start with both the system 
and its environments being made of fermions. Let $a_i, a^\dag_i$ and $b_{\alpha k}, b^\dag_{\alpha k}$ denote the 
creation and annihilation operators of fermions of the system and the environment $\alpha$, respectively, 
which satisfy the standard fermionic anti-commutation 
relationships. Then the most basic process for the underlying matter exchange between the system and 
its environments can be described effectively through the following Hamiltonian:
\begin{align}
H(t) = &\sum_i \varepsilon_{ij} (t) a^\dag_i a_j + \sum_{\alpha k} \epsilon_{\alpha k}(t)
 b^\dag_{\alpha k} b_{\alpha k} 
 +\sum_{\alpha ik}\Big(V_{i \alpha k}(t) a^\dag_i b_{\alpha k}
 + V^*_{i \alpha k}(t)  b^\dag_{\alpha k} a_i\Big)  ,  \label{FAH}
\end{align}
where the system-environment coupling, i.e. the last term in the above equation, is realized through 
particle tunnelings between the system and its environments. The index $\alpha$ denotes different
environments in contact with the same systems. I let all the parameters in the Hamiltonian be time-dependent
because current nanotechnologies are capable to tune these parameters effectively through various 
external fields, such as bias and gate voltages in nanoelectronics. The form of Eq.~(\ref{FAH}) is indeed
a generalized Fano-Anderson Hamiltonian in condensed matter physics \cite{Mahan2000},
as it was originally introduced by Anderson \cite{Anderson1961} and Fano \cite{Fano1961} independently, 
to describe impurity electrons coupled to continuous states in solid-state physics, and discrete states 
embedded in a continuum in atomic spectra, respectively.  In the past two decades, one has started with
such Hamiltonians to investigate various electron transport phenomena and decoherence dynamics in 
semiconductor nanoelectronics and spintronics for mesoscopic physics, where the nano-scale electron system 
is in contact with two (or more) electrodes which serve as the electronic reservoirs
\cite{Hang1996,Imry2002}.  

The system interacting with its environments in terms of the
Fano-Anderson Hamiltonian (\ref{FAH}) has also been applied to the case where both the system and the environments 
are made of bosons as composite particles of fermions, such as atoms that Fano originally proposed 
\cite{Fano1961}, or more fundamentally made of scalar fields \cite{Lee1954}.  Correspondingly, 
the creation and annihilation operators, $a_i, a^\dag_i$ and $b_{\alpha k}, b^\dag_{\alpha k}$, would then obey bosonic 
commutation relationships. The generalized Fano-Anderson Hamiltonian was also 
refereed to the multi-level Lee-Friedrich Hamiltonian \cite{Friedrichs1948,Lee1954,Prigogine1992} which models
the processes of the quantum mechanical resonances and decays of the discrete state or localized particles with a 
continuum in different contexts of atomic and molecular physics, nuclear physics and quantum field theory 
\cite{Lee1954,Tannoudji92,Knight1990}.
If the system and its environment are made of massless particles, such as photons, the Hamiltonian 
of Eq.~(\ref{FAH}) would describe the photon scattering processes, corresponding to photon
tunnelings between the system and the environment, such as photon loss into the free space (spontaneous 
emissions) in quantum optics as well as in the various new development of nanophotonics 
\cite{Carmichael1993,Lam00,ANNP2012}.  

The next order contribution to the exchanges of matters and energies between the system and its 
environments originates from particle-particle interactions. Correspondingly, a coupling 
Hamiltonian arisen from particle-particle interactions between the system and its environments should 
be added to Eq.~(\ref{FAH}),
\begin{align}
H'_{SE}=\!\!\! \sum_{\alpha, ijkl}\!\! \! \Big(\!V^{(1)}_{\alpha,ijkl} a^\dag_i 
a^\dag_j b_{\alpha l} b_{\alpha k} + V^{(2)}_{\alpha,ijkl}
a^\dag_i b^\dag_{\alpha l} b_{\alpha k} a_j + {\rm H.c.}\!\Big) , \label{HT2}
\end{align}
where the first term plus its Hermitian conjugate correspond to particle-pair exchanges between 
the system and its environment $\alpha$, and the second term only involves energy exchanges 
(particle-particle scattering interactions) between them.   Including the interacting Hamiltonian of 
Eq.~(\ref{HT2}) makes the system and its environments become a typical interacting 
many-body problem \cite{Mahan2000}. Except for  weak particle-particle interactions, where the perturbation 
approach can be applied, the problem associated with Eq.~(\ref{HT2}) cannot be solved exactly in general, 
just like the strongly-correlated electronic systems in condensed matter physics and the low-energy 
quantum chromodynamics (QCD) in particle physics. Related to exact master equation and 
non-Markovian dynamics, I have developed a general transport theory using the 
closed-time path integral approach and the associated quantum  Boltzmann equation in terms 
of loop expansions technique within the quantum field theory framework two decades ago
\cite{Zhang1992},  but this will be not the main approach I will discuss in this article. 

The above discussions focus on open quantum systems where both the system and its environments 
are made of the same type of particles, either fermions or bosons. However, we also often have the 
situation where the system is a fermionic system and its environments are made of bosons.  A typical 
example of this type is the system coupled to its environment through matter-light interaction \cite{Tannoudji92}, 
where the system-environment coupling is determined by electron-photon interactions under the 
fundamental theory of quantum electrodynamics (QED),
\begin{align}
H_{e-p}&= -e\!\! \int \!\! d^3x \psi^\dag(x) \gamma_\mu \psi(x) A^\mu(x) 
= \sum_{pq} \!\Big(\! V_{pq}a^\dag_{p+q} a_{p}b_q + 
 V^*_{pq}a^\dag_{p} a_{p+q}b^\dag_{q}\Big) ,   \label{QED}
\end{align}
where $\psi(x)$ is the electron field, $A^\mu(x)$ is the vector potential of the electromagnetic field, 
$e$ is the electron charge, and $\gamma_\mu$ are the Dirac $\gamma$-matrices \cite{Peskin1995}. 
In the second equality of Eq.~(\ref{QED}), I have ignored the antiparticle component (positron) 
of the electron field because it usually has no contribution in the nonrelativistic regime.  
The electron-phonon coupling between the system and its environment in condensed 
matter physics shares the same coupling Hamiltonian. The dynamics of the system with such 
electron-photon interaction has also been extensively studied in the framework of perturbation 
expansions. A two-level system coupled to an electromagnetic field  in quantum optics, 
i.e. several different kind of spin-bosom models that I will discuss in Sec. III, can be in principle 
reduced from (\ref{QED}). 
The systematic nonperturbation approach I developed in \cite{Zhang1992}  can also 
be applied to the systems of Eq.~(\ref{QED}) but this is not the exact master 
equation approach that I will discuss in this article.

As a result of the above analysis, I will concentrate on the dynamics of open quantum systems 
with the generalized Fano-Anderson Hamiltonian or multi-level Lee-Friedrichs Hamiltonian, 
given explicitly by Eq.~(\ref{FAH}). This Hamiltonian describes the underlying 
exchanges of matters and/or energies between the system and its environments on one hand, and on the 
other hand, is applicable to tremendous nano-scale quantum devices in the investigation of 
nanoelectronics and nanophotonics, including the new emerging research field of topological 
phase of matter. 

\section{Exact master equation of open quantum systems}
\label{sec:1}
Now, I begin with discussion about how to derive the exact master equation of the open quantum systems described 
by the Hamiltonian of Eq.~(\ref{FAH}), and then review probably all known exact master equations in the
literature. Let us start with the basic assumption \cite{Fey1963,Leggett1987} 
that the system and all different environments $\{ \alpha \}$ are initially decoupled. Extension to the initially
entangled state between the system and the environment has also been partially worked out and I will 
discussed it later \cite{Tan2011,PRB2015,Huang18}.   
More specifically, let the system be in an arbitrary initial state $\rho(t_0)$, and environment be 
initially in a thermal state, 
\begin{align}
\rho_{\rm tot}(t_0)= \rho(t_0) \otimes \rho_E(t_0), ~~~ \rho_E(t_0)= \frac{1}{Z}e^{-\sum_\alpha \beta_\alpha H_{\alpha E}} . \label{ids}
\end{align}
Here $Z= \prod_{\alpha k} (1 \mp e^{-\beta_\alpha (\varepsilon_{\alpha k}-\mu_\alpha)})^{\mp 1}$ is the partition 
function of the environments. Different environment $\alpha$ could have different
chemical potential $\mu_\alpha$ and different temperature $\beta_\alpha =1/k_B T_\alpha$. The up and down signs 
of $\mp$ correspond to the environment being bosonic and fermionic systems, respectively.

As I have discussed in the Introduction, if the initial state of the total system (the system plus its environments) is not in
a pure state, dynamics of the total system is governed  by  von Neumann equation, which is still a {\it unitary evolution 
equation of motion},
\begin{align}
i\hbar \frac{d\rho_{\rm tot}(t)}{dt}= [H(t), \rho_{\rm tot}(t)]. 
\end{align}  
The exact time-evolving reduced density matrix of the open system, 
defined by $\rho(t)={\rm Tr}_E [\rho_{\rm tot}(t)]$, can be obtained after taking the trace 
over all the environment states,
\begin{align}
\rho(t)= {\rm Tr}_E [{\cal U}(t,t_0)\rho(t_0)\! \otimes \!\rho_E(t_0){\cal U}^\dag(t,t_0)] ,
\end{align}
where ${\cal U}(t,t_0)= {\cal T}\exp\{\!- i \!\int^t_{t_0}\!d\tau H(\tau)\}$ is the unitary evolution operator of the 
total system, ${\cal T}$ is the time-ordering operator, and $H(\tau)$ is the Hamiltonian of Eq.~(\ref{FAH}).
As we will see later, after taken the trace over all the environment states, the evolution of the reduced density
matrix becomes in general non-unitary.

Without taking an approximation, one can take trace over the environment states by integrating out 
all the environmental degrees of freedom, using the Feynman-Vernon influence functional.  
However, because the system and environments are
made of either fermionic or bosonic particles, one must extend the path-integral into the coherent-state
representations, and in particular special attention must be paid on the initial and final boundary conditions 
of path-integrals for both fermion and boson variables, and also on additional antisymmetric property in 
closed-loop integrals for fermionic (Grassmann) variables \cite{PRB2008,NJP2010,ANNP2012}.
It should also note that Feynman-Vernon influence functional makes the influence of the environment
into an effective action on the system. It is still far to solve the dynamics of open systems by only integrating 
out all the environmental degrees of freedom through the influence functional approach.   

In fact, after a cumbersome calculation of the propagating function for the reduced density matrix 
in coherent-state path-integral representation, and a consummate elimination of the initial variable dependence
of all paths through a nontrivial transformation (see explicitly, Eq.~(29) in Ref.~\cite{PRB2008} 
or Eq.~(A1) in Appendix A of \cite{NJP2010} for fermion systems and Eq.~(27) in Ref.~\cite{ANNP2012}
for boson systems), we obtain the following exact master equation \cite{PRB2008,NJP2010,ANNP2012}:
\begin{align}
\dot{\rho}(t)   = -i\big[  H'_{S}(t,t_0),  \rho(t)\big] 
+\sum_{ij}\Big\{ \gamma_{ij}(t,t_0)
\big[2a_{j}\rho(t) a_{i}^{\dagger}  -a_{i}^{\dagger}a_{j}\rho(t)  -\rho(t)
a_{i}^{\dagger}a_{j}\big] ~~~~ \nonumber \\
 +  \widetilde{\gamma}_{ij}(t,t_0) \big [a_{i}%
^{\dagger}\rho(t) a_{j} \, {\pm} \, a_{j}\rho(t)  a_{i}^{\dagger}\mp a_{i}^{\dagger}a_{j}%
\rho(t)  - \rho(t) a_{j}a_{i}^{\dagger}\big]\Big\},
\label{Exact-ME}%
\end{align}
where  the $+/-$ signs correspond to boson/fermion systems.
The first term in Eq.~(\ref{Exact-ME}) provides still a unitary evolution of system with the 
environment-induced renormalized Hamiltonian 
$H_{S}'(t,t_0)  =\sum_{ij} {\varepsilon^{\prime}_{i j}} (t,t_0)a_{i}^{\dagger}a_{j}$.
The second and third terms result in non-unitary evolutions for the
environment-induced dissipation and fluctuations, respectively. 
This exact master equation has nonperturbatively taken into account all the
environment-induced back-reactions up to all orders, which are embedded into 
the renormalized energy matrix of the system $\varepsilon_{ij}(t,t_0)$,
the dissipation coefficients $\gamma_{ij}(t,t_0)$ and the fluctuation coefficients 
$\widetilde{\gamma}_{ij}(t,t_0)$.

We found that all these time-dependent coefficients in the master equation
are determined exactly by the nonequilibrium
Green's functions through the following relations \cite{PRB2008,NJP2010,ANNP2012},
\begin{subequations}
\label{emec}
\begin{align}
& \varepsilon_{ij}^{\prime}(t,t_0)  =\frac{i}{2}\left[
\boldsymbol{\dot{u}}\left(  t,t_{0}\right)  \boldsymbol{u}^{-1}\left(
t,t_{0}\right)  -\text{H.c.}\right]_{ij}  ,  \label{er} \\
& \gamma_{ij}(t,t_0)  =-\frac{1}{2}\left[  \boldsymbol{\dot
{u}}\left(  t,t_{0}\right)  \boldsymbol{u}^{-1}\left(  t,t_{0}\right)
+\text{H.c.}\right] _{ij} ,   \label{dssc} \\
& \widetilde{\gamma}_{ij}(t,t_0)  =\boldsymbol{\dot{v}}_{ij}%
\left(  t,t\right)  -\left[  \boldsymbol{\dot{u}}\left(  t,t_{0}\right)
\boldsymbol{u}^{-1}\left(  t,t_{0}\right)  \boldsymbol{v}\left(  t,t\right)
+\text{H.c.}\right]_{ij} . \label{gamma-gmmatilde}%
\end{align}
\end{subequations}
The functions  $\boldsymbol{u}(t,t_{0})$ and $\boldsymbol{v}(t,t)$ are the nonequilibrium dissipative 
propagating Green function (which may be simply called as the spectrum Green function) and fluctuated (correlation 
or Keldysh) Green function (which is related to the lesser Green function, see more details in Sec.~4), 
respectively. They obey the following integro-differential equations%
\begin{subequations}
\label{uvt}
\begin{align}
&\frac{d}{d\tau}\boldsymbol{u}(\tau,t_{0})
+i\boldsymbol{\varepsilon}(\tau) \boldsymbol{u}(\tau,t_{0})  +\!\!\int 
_{t_{0}}^{\tau}\!\! \!\! d\tau^{\prime}  \boldsymbol{g}(\tau,\tau^{\prime})
\boldsymbol{u}(\tau^{\prime},t_{0})  =0, \label{ute} \\
&\frac{d}{d\tau}\boldsymbol{v}(\tau,t)  +i\boldsymbol{\varepsilon}(\tau) 
\boldsymbol{v}(\tau,t)  +  \!\!\int_{t_{0}}^{\tau} \!\!\!\! d\tau^{\prime
}\boldsymbol{g}(\tau,\tau^{\prime})  \boldsymbol{v}(\tau^{\prime},t) 
=\int_{t_{0}}^{t}\!\! d\tau^{\prime}\widetilde{\boldsymbol{g}}(\tau
,\tau^{\prime})  \boldsymbol{u}^{\dagger}(\tau^{\prime},t_{0})     \label{vte}
\end{align}
\end{subequations}
subjected to the boundary conditions $\boldsymbol{u}(t_{0},t_{0})  =\mathbf{1}$ 
and $\boldsymbol{v}(t_0,t)=0$ with $\tau\in\left[  t_{0},t\right]  $. The energy matrix
$\boldsymbol{\varepsilon}(\tau) \equiv \{ \, \varepsilon_{ij} (\tau)\, \}$ is the $N\!\times\! N$ 
single-particle energy matrix of the system. The time non-local dissipation and fluctuation kennels in the 
above integro-differential equations, $\boldsymbol{g}\left(  \tau
,\tau^{\prime}\right) $ and $\widetilde{\boldsymbol{g}}\left(\tau,\tau^{\prime}\right)$, are given by 
\begin{subequations}
\begin{align}
&\boldsymbol{g}_{ij}( \tau,\tau^{\prime})  =\sum_{\alpha k} V_{i\alpha k}(\tau')V^*_{j\alpha k}(\tau)
\exp\!\Big\{\!-i\! \!\int_\tau^{\tau'} \!\!\!\!d\tau_1\epsilon_{\alpha k} (\tau_1)\Big\} ,   \label{ik} \\
&\widetilde{\boldsymbol{g}}_{ij}(\tau,\tau')  =\sum_{\alpha k}V_{i\alpha k}(\tau')V^*_{j\alpha k} (\tau)
\big\langle b^\dag_{\alpha k}(t_0) b_{\alpha k}(t_0) \big\rangle_E
\exp\!\Big\{\!-i\! \!\int_\tau^{\tau'} \!\!\!\!d\tau_1\epsilon_{\alpha k} (\tau_1)\Big\},  \label{gtt}
\end{align}
\end{subequations}
The initial environment correlation function,
$\big\langle b^\dag_{\alpha k}(t_0) b_{\alpha k}(t_0) \big\rangle_E = f (  \epsilon_{\alpha k},T_\alpha)$, 
which is the initial particle distribution of the bosons or fermions in 
environment $\alpha$ with the chemical potential $\mu_\alpha$ and the temperature 
$T_\alpha$ at time $t=t_0$, i.e., $f(\epsilon,T_\alpha)=$ $[e^{(\epsilon
-\mu_\alpha)/k_{B}T_\alpha} {\mp} 1 ]^{-1} $. 
In the case the energy spectra of the environments and the system-environment
couplings are all time-independent, the time non-local dissipation and fluctuation kennels
are reduced to 
$\boldsymbol{g}( \tau,\tau^{\prime})  =  \int\frac{d\epsilon}{2\pi}\boldsymbol{J}(\epsilon)  
e^{-i\epsilon (\tau-\tau')} $, and
$\widetilde{\boldsymbol{g}}_{ij}(\tau,\tau') =  \int\frac{d\epsilon}{2\pi}\boldsymbol{J}(\epsilon) 
f(\epsilon,T_\alpha)  e^{-i\epsilon (\tau-\tau')}$, where  $\boldsymbol{J}(\epsilon)$ 
is the spectral density of the open system,
\begin{align}
\boldsymbol{J}_{ij}(\epsilon) \equiv 2\pi \sum_{\alpha k} V_{i\alpha k}V^*_{j\alpha k}
\delta(\epsilon_{\alpha k} - \epsilon).  \label{dsd}
\end{align}

As one note that the exact master equation (\ref{Exact-ME}) is time-convolutionless. 
Because this exact master equation can unambiguously capture non-Markovian dynamics 
of open system, the key point in this theory is that the dissipation coefficients $\gamma(t,t_0)$ and 
fluctuation coefficients $\widetilde{\gamma}(t,t_0)$ in the master equation are determined 
by the nonequilibrium Green functions $\boldsymbol{u}(t,t_{0})$ and $\boldsymbol{v}(t,t)$ 
which obey the time-convolution equations of motion. It is this time-convolution 
structure embedded in its dissipation and fluctuation coefficients makes the time-convolutionless 
master equation (\ref{Exact-ME}) become capable to depict precisely all the possible non-Markovian memory effects. 
It is indeed also a necessary condition that any quantum theory of open systems that can capture 
non-Markovian dynamics must involve time-convolution equation of motion in certain manners. 
Thus, it is straightforward to conclude that the Born-Markovian type master equation that has 
widely used in the literature can only describe memoryless Markov processes, because it 
is time-convolutionless and it also does not involve any time-convolution structure in the dissipation 
and fluctuation dynamics.   

It may be also worth pointing out that the exact master equation (\ref{Exact-ME}) can be further
rewritten in terms of the Lindblad operator form, as we have shown in the previous work 
\cite{ANNP2012} 
\begin{subequations}
\label{eme0}
\begin{align}
\frac{d\rho(t)}{dt} = & \frac{1}{i}[\tilde{H}_S(t,t_0),
\rho(t)]  + \sum_{ij}\tilde{\boldsymbol{\gamma}}_{ij}(t,t_0)
L_{a^\dag_i,a_j}[\rho(t)] ~~~~ \notag \\ 
&+ \sum_{ij}[2 \boldsymbol{\gamma}_{ij}(t,t_0) \pm
\widetilde{\boldsymbol{\gamma}}_{ij}(t,t_0)]L_{a_j,a^\dag_i}[\rho(t)] \
, \label{eme} 
\end{align}
where the super-operator  
$L_{a_i,a^\dag_j} [\rho(t)] $ is defined as the standard Lindblad operator
\begin{align} 
L_{a_i,a^\dag_j} [\rho(t)] \equiv
a_i\rho(t)a^{\dag}_j - \frac{1}{2}a^{\dag}_ja_i\rho(t) -
\frac{1}{2}\rho(t) a^{\dag}_ja_i \, .
\end{align}
\end{subequations}
However, I must emphasize that the above exact master equation in the Lindblad form is completely 
different from the usual Lindblad semigroup master equation derived from semigroup mapping 
\cite{GKS,Lindblad}. The Lindblad semigroup master equation neglects all reservoir memory 
effects (therefore, it does not involve any time-convolution structure) so that it is only applicable 
to Markov dynamics.  Here, the master equation (\ref{Exact-ME}), although it can be written formally in a Lindblad 
form, i.e.~Eq.~(\ref{eme}), can capture precisely the non-Markovian dynamics, as I have 
emphasized,  because it is exact and because its dissipation and fluctuation coefficients 
are determined by time-convolution equations
of motion.  It may also be interest to notice that the Lindblad form of the exact master equation 
(\ref{eme}) mixes the dissipation and fluctuation in the last two non-unitary terms such that the 
fluctuation-dissipation theorem is not manifested.  In the form of the exact master equation 
(\ref{Exact-ME}), the last two non-unitary terms describes the dissipation and fluctuations explicitly 
and separately, and the dissipation and fluctuation dynamics are connected remarkably by the 
intrinsic nonequilibrium fluctuation-dissipation theorem in the time domain, see the discussion 
later. In this sense, the Lindblad form of the master equation used widely in the literature is physically not 
so essential in understanding the dissipation and fluctuation dynamics in open quantum systems. 

As it has also been pointed out in Sec. II, the earliest master equation derived from the original 
Feynman-Vernon influence functional \cite{Fey1963} was obtained by Caldeira and 
Leggett for the quantum Brownian particle in an Ohmic environment
\cite{Caldeira1983}.  An extended exact master equation with general color noise was 
derived later By Haake and Reibold \cite{QBM2} using the equation of motion approach, 
and then is reproduced by Hu {\it et al.} \cite{HPZ1992} 
with the Feynman-Vernon influence functional approach again. The further extension to the exact master equation 
with more general initial states is also found by Grabert {\it et al.} \cite{Grabert88,Grabert1997}. 
The exact master equation for the CL Hamiltonian 
in \cite{Caldeira1983,QBM2,HPZ1992,Grabert1997} looks quite different from the exact master equation (\ref{Exact-ME}) 
we obtained from Eq.~(\ref{FAH}), not only on the different formulation (the former cannot
have a Lindblad form and looks complicated) but also on the different physics given by these two Hamiltonians. 
Specifically, in the system-environment Hamiltonian of the CL model,  
$H_{\rm SE}=\sum_k c_kx q_k=\sum_k c'_k(a^\dag b_k + a b^\dag_k + a^\dag b_k^\dag + a b_k)$,
the last two terms correspond to the processes of generating or annihilating two quanta of 
energy out of the blue, from {\it nothing}, which is quantum mechanically unreliable \cite{intep}. 
One should be aware that the CL model is a semi-empirical model.
Its physical motivation is to derive the classical dissipation motion of a Brownian particle 
from quantum mechanics, where the 
contribution from the last two terms is negligible.  
Thus removing these paring terms in the CL
Hamiltonian will reduce the CL model into the Fano-Anderson type model of Eq.~(\ref{FAH}). In this situation,
the physics described by the master equation for quantum Brownian motion can also be 
covered by the master equation of Eq.~(\ref{Exact-ME}). 

There are also many attempts to derive the exact master equation for other open quantum 
systems, in particular, the spin-boson model \cite{Leggett1987}:  $H = \frac{\varepsilon}{2}\sigma_+\sigma_- + \frac{\Delta}{2}\sigma_x
+ \sum_k \omega_k b^\dag_kb_k+ \sigma_x \sum_k V_k(b_k + b^\dag_k)$. The spin-boson model is indeed a particular 
realization of the CL model with the particle moving in a double-well potential and coupling 
to a thermal both, and only considers the particle hops between the double wells rather 
a continuous variable \cite{Leggett1987}. 
One can use the Feynman-Vernon influence functional method to completely integrate out the bosonic bath 
degrees of freedom, resulting in a closed form of the effective action for spin dissipation 
\cite{Leggett1987}. However, because of the 
non-commuting spin operators, 
the spin dissipation then contains heavy nonlinearity and becomes not exactly solvable, so does 
have not a closed exact master equation be found so far \cite{Leggett1987}.  
Also, similar to the original CL model, the spin-boson model also contains quantum 
mechanically unreliable processes \cite{intep}, i.e. spin can simultaneously emit photons when it is excited 
from the ground state to the excited state. Removing these unreliable processes (corresponding 
to the so-called rotating-wave approximation in the literature),  the spin-boson model is reduced to the
multimode Jaynes-Cummings (JC) model \cite{JCM1963}: $H = \frac{\varepsilon}{2}\sigma_+\sigma_- 
+ \frac{\Delta}{2}\sigma_x + \sum_k \omega_k b^\dag_kb_k+  \sum_k V_k(\sigma_+b_k + \sigma_-b^\dag_k)$,
where the term $\frac{\Delta}{2}\sigma_x$ also plays the role of an external deriving field. Even for this 
simplified spin-boson model, no exact master equation has been found.

There are some attempts to derive the exact master equation for this multimode JC model \cite{AH2000,Shen2014},
using the Feynman-Vernon influence-functional method, in which they treat the spin operators as fermion 
operators and then apply the spin coherent state with Grassmann variable to perform the spin path integral. 
This treatment is mathematically and physically incorrect, as it also has been challenged recently \cite{Whalen2016}. 
The incorrectness comes from the fact that spin does not obey 
fermion statistics, and the Grassmann variables are introduced only for fermions because of their anticommutation 
property \cite{Faddeev1980}. Spin coherent state is defined on a sphere with continuous variables ($ 0\le \theta\le \pi,
0\le \phi\le 2\pi$) \cite{Gilmore1974,Zhang1990}. Write the spin coherent state with Grassmann variables 
\cite{AH2000,Shen2014} only covers a small subspace of the spin Hilbert space such that it excludes
the nonlinearity of spin dynamics induced by the thermal bath (or by the driving field). This is why the results in 
\cite{AH2000,Shen2014} are incorrect.  In fact, the correct expression of spin operators in terms of
fermion operators is $\vec{\boldsymbol S}= \frac{1}{2}a^\dag_i \vec{\boldsymbol \sigma}_{ij} a_j$.  Then the multimode 
JC model has the QED Hamiltonian form of Eq.~(\ref{QED}). After integrated out the bosonic
bath degrees of freedom, the resulting effective action for the spin dynamics contains 
time non-local  fermion-fermion interactions that are more complicated than the interactions in
Eq.~(\ref{HT2}) and are not solvable in general. This is why a closed form 
of the exact master equation for spin-boson systems is so difficult to be derived. 

But there is an exception when the spin is
initially in the excited state and couples to the bath at zero temperature and no driving field ($\Delta=0$), then the dynamics
of spin involves only one photon, and the nonlinearity of spin dynamics does not manifest. Only in this
special case, the result in \cite{AH2000,Shen2014} is accidentally correct due to the lack of spin nonlinearity.
The resulting master equation is 
\begin{align}
\frac{d\rho(t)}{dt} = -i[\varepsilon'(t,t_0)s_+s_-, \rho(t)] + \gamma(t,t_0)\{\sigma_- \rho(t) \sigma_+ - \sigma_+ \sigma_-\rho(t)
- \rho(t)\sigma_+ \sigma_- \} .   \label{spinme}
\end{align}
This exact master equation can even be easily derived from Schr\"{o}dinger equation \cite{Garraway1997,Breuer1999}.
The spin energy renormalization $\varepsilon'(t,t_0)$ and the dissipation coefficient  $\gamma(t,t_0)$ are actually determined 
by the same equation Eq.~(\ref{emec}).  In other words, the master equation
(\ref{spinme}) is indeed a special case of  Eq.~(\ref{Exact-ME}) with the bosonic bath at zero temperature.
For master equation Eq.~(\ref{Exact-ME}) at zero temperature, $v(t,t)=0$ [see the solution of Eq.~(\ref{v(t)}) with (\ref{gtt})] so that 
the fluctuation coefficient vanishes: $\widetilde{\gamma}(t,t_0)=0$.

The remaining open quantum systems that can be solved exactly in terms of exact master equation 
is the pure dephasing models, in which the system-bath coupling commutes with the system 
Hamiltonian. A typical pure dephasing model is a spin coupled to a bosonic thermal bath in such a way that
$H= \varepsilon \sigma_z + \sum_k \omega_k b^\dag_k b_k + \sigma_z \sum_k V_k (b_k + b^\dag_k)$.  
The corresponding exact master equation is \cite{Doll2008,Goan2010},
\begin{align}
\frac{d\rho(t)}{dt} = -i[\varepsilon \sigma_z, \rho(t)] + \gamma^{(2)}(t,t_0)\{\sigma_z \rho(t) \sigma_z - \rho(t) \},
\end{align}
where the dephasing coefficient $\gamma^{(2)}(t,t_0)\!=\! 2 \int \!d\omega  J(\omega) 
\coth \!\big(\frac{\omega}{2k_BT}\!\big) \! \cos(\omega (t-t_0))$,
which corresponds to a second-order perturbation result of $\gamma(t,t_0)$ from Eq.~(\ref{dssc}),
and therefore does not contain non-Markovian 
dynamics. This makes the exact master equation identical with the Markovian master equation,  which
is an exception only for the spin with dephasing noise from a bosonic thermal bath \cite{Doll2008}.
Very recently, we study the local gate-control-induced change fluctuations to Majorana zero modes in topological 
 quantum computing \cite{Schmidt2012} which corresponds to a new kind of pure dephasing noise to the Majorana zero 
 modes in a fermion bath
 \cite{PRB18}, $H= \sum_k \varepsilon_k b^\dag_k b_k + \sum_k V_k(b_k+b^\dag_k)\lambda$, where $\lambda$ 
 is the Majorana quasiparticle operator, $\lambda^\dag=\lambda$, i.e. the particle is also its own antiparticle, and 
$\{ b_k \}$ represent the fermion bath. The Hamiltonian for Majorana zero modes is zero. The exact 
master equation is \cite{PRB18}
 \begin{align}
\frac{d\rho(t)}{dt} = \gamma(t,t_0)\{\lambda \rho(t) \lambda - \rho(t) \}   \label{meme}
\end{align}
which cannot be obtained from the second-order perturbation, where the damping coefficient $ \gamma(t,t_0)$ in 
Eq.~(\ref{meme}) is  still determined by the same equation Eq.~(\ref{dssc}) (up to all orders)
with a slight change to the integral kernel in Eq.~(\ref{ute}), $g(\tau,\tau')\rightarrow g(\tau,\tau')+g(\tau',\tau)$,
due to the particle-hole symmetry.  Therefore, the Majorana decoherence  can be very non-Markovian. 
 
There are also other generalized master equations, such as the time-non-local Nakajawa-Zeneng master equation
\cite{Nakajima,Zwanzig}, the time-convolution-less expansion of the master equation \cite{Breuer1999,Breuerbook}, 
the master equation with hierarchical expansion \cite{Jin2008}, and the generalized master equation for
$N$-level-bosom models \cite{Thorwart2001}, etc., and although these master equation can be formally 
exact, they are either intractable or expressed in terms of infinite series expansions. Further  
approximations must be made in solving these master equations for practical applications, and therefore
cannot be considered as exact in practice.  In conclusion, Eq.~(\ref{Exact-ME}) is the most general exact 
master equation of open quantum systems that one can 
analytically solve, from which the non-Markovian dynamics of open quantum systems 
can be universally investigated by simply solving the equation of the Green function 
Eq.~(\ref{uvt}) \cite{PRL2012}, as I will discuss in the next section. 

\section{General non-Markovian dynamics}
\label{sec:2}
Once we have the exact master equation that can precisely capture non-Markovian dynamics,
we are capable to discuss the general properties of non-Markovian dynamics in open quantum 
systems. As one can see from the last Section, the dissipation and fluctuation coefficients in 
the exact master equation are completely determined by the nonequilibrium Green functions of Eq.~(\ref{uvt}). 
Physically, it may be more transparent to express these nonequilibrium Green functions in terms of 
field operators.  After eliminated the environment degrees of freedom from Heisenberg equation
of motion, we have the exact 
quantum Langevin equation from Eq.~(\ref{FAH}) \cite{Tan2011,Yang2014,PRB2015}, 
\begin{align}
\frac{d a_i(t)}{dt}= \sum_{j} \Big\{-i \varepsilon_{ij} a_j(t) - \!\! \int^t_{t_0} \!\! d\tau 
\boldsymbol{g}_{ij}(t,\tau)a_j(\tau) \Big\} + f_i(t)
\end{align}
where $\boldsymbol{g}(t,\tau)$ is given by Eq.~(\ref{ik}), and $f_i(t)= -i\sum_{\alpha k} 
\!V_{i \alpha k}(t) b_{\alpha k}(t_0) e^{-i \int^t_{t_0}\! \epsilon_{\alpha k}(\tau)d\tau}$ 
is the environment-induced noise operator that depends explicitly on the system-environment 
coupling $V_{i\alpha k}(t)$ at the time $t$ and on the initial state of the environment through
the operator $b_{\alpha k}(t_0)$. 

The linearity of the above quantum Langevin equation leads to the general solution 
\begin{align}
a_i(t)= \boldsymbol{u}_{ij}(t,t_0) a_j(t_0) + F_i(t)~,~~{\rm and} ~~~ 
F_i(t)= \sum_j \int^t_{t_0} \!d\tau \boldsymbol{u}_{ij}(t,t_0)f_j(\tau) ,
\end{align}
where  $\boldsymbol{u}_{ij}(t,t_0) \equiv \langle [a(t), a^\dag(t_0)]_{\mp} \rangle $ is the propagating (retarded) 
Green function of Eq.~(\ref{ute}), and   $\boldsymbol{v}_{ij}(\tau,t)= \langle F^\dag_j(t) F_i(\tau)\rangle$ 
is the correlation Green function of Eq.~(\ref{vte}) that manifests the nonequilibrium fluctuation-dissipation
theorem.  These solutions clearly show how the propagating and correlation Green functions describe completely
the dissipation and fluctuation dynamics of open systems, and why the dissipation and fluctuation 
coefficients are determined by these nonequilibrium Green functions.  
Moreover, the above results are obtained without the assumption of the initial decoupled states, 
namely, {\it they can be applied to the cases with initial entangled states between the system 
and environments} \cite{Tan2011,PRB2015,Yang2016b}.  When the initial states of system and 
environment are decoupled as that given by Eq.~(\ref{ids}), we can easily obtain the lesser Green function 
in Keldysh's nonequilibrium Green function technique \cite{Schwinger1961,Keldysh1965,Kadanoff1962},
which is defined by $-iG^<_{ij}(\tau, t) \equiv \langle a^{\dag}_j(t) a_i(\tau)  \rangle=\boldsymbol{u}_{il'}
(\tau,t_0) \langle a^{\dagger}_l (t_0) a_{l'}(t_0)\rangle \boldsymbol{u}^\dag_{jl}(t,t_0) 
  + \boldsymbol{v}_{ij}(\tau,t)$. Thus, these results give the whole physical picture of dissipation 
and fluctuation dynamics in terms of nondequilibrium Green functions in open quantum systems through 
the exact master equation formalism of Sec.~III.  
As a result, the general non-Markovian dynamics of open quantum systems can be fully manifested 
in terms of nonequilibrium Green functions. 

Explicitly, the dissipation dynamics of open systems is determined by 
its nonequilibrium propagating Green function $\boldsymbol{u}(t,t_0)$, its general solution consists of 
nonexponential decays and dissipationless oscillations \cite{PRL2012}. For simplicity, we consider 
the open system of a single particle in the state with energy $\varepsilon_s$, in contact with a reservoir.
Then the general solution of the propagating Green function of Eq.~(\ref{ute}) has the form
\begin{align}
u(t,t_0)  =\sum_{j}\mathcal{Z}_{b_{j}}e^{-i\varepsilon_{b_{j}}(t-t_{0}) }+\!\! \int \!\! 
d\epsilon \mathcal{D}_d(\epsilon)  e^{-i\epsilon( t-t_{0})  } . \label{u(t)}%
\end{align}
Here, the first term contains dissipationless oscillations, as the contribution of localized 
bound states in the system, after taking into account the effect from the environment. 
The corresponding localized bound state energy (frequency) $\varepsilon_{b_{j}}$ and its 
amplitude $\mathcal{Z}_{b_{j}}$ are determined by the pole of the propagating Green function
in the Fourier or Laplace transform: 
\begin{align} \varepsilon_{b _{j}} - \varepsilon_{s} - \Delta( \varepsilon _{b_{j}} ) = 0~, ~~{\rm and}
~~~\mathcal{Z}_{b_{j}} = \frac{1}{ 1 - \partial_{\varepsilon} \Sigma (\varepsilon) \vert_{\varepsilon = \varepsilon _{b_{j}}}} ,
\end{align} 
where $ \Sigma (\varepsilon)\!= \!\int \frac{d \epsilon'}{2\pi} \frac{J (\epsilon')}{\varepsilon - \epsilon'}$ 
is the environment-induced self-energy correction to the system energy [the Fourier transform 
of the dissipation kernel in Eq.~(\ref{uvt})], and  $\Delta (\varepsilon) 
= {\cal P}\!\int \! \frac{d\epsilon'}{2\pi} \frac{J(\epsilon')}{\varepsilon-\epsilon'}$ is the principal value 
of the integral. 
The second term in Eq.~(\ref{u(t)}) is the continuous part of the spectrum of the single particle moving in the environment,
which leads to the dissipative dynamics (and it is in general a nonexponential damping or decay) of the system. 
The corresponding dissipation spectrum $\mathcal{D}_d(\epsilon)$ is given by 
\begin{align}
\mathcal{D}_d (\epsilon) = \frac{J(\epsilon)}{[\epsilon - \epsilon_s - \Delta (\epsilon)]^{2} + \pi ^{2} 
J ^{2} (\epsilon)},  \label{ds}
\end{align}
It is particularly important to note that the localized bound states [the first term in Eq.~(\ref{u(t)})] 
only exist if the spectral density $J(\epsilon)$ contains band gap(s) or zero energy points with 
sharp slopes, while the dissipation spectrum $\mathcal{D}_d (\epsilon)$ [proportional to 
$J(\epsilon)$, see Eq.~(\ref{ds})] is crucially determined by the spectral density profile, 
as we have emphasized in \cite{PRL2012}.  In the steady-state limit $t=t_s  \rightarrow  \infty$, 
only the dissipationless oscillation terms remain,  $u(t_s,t_0)  =\sum_{j}\mathcal{Z}_{b_{j}}
e^{-i\varepsilon_{b_{j}}(t_s-t_{0})}$.  If there is no localized bound state, $u(t_s,t_0) \rightarrow 0$
in the steady-state limit, corresponding to a complete relaxation process.

On the other hand, the general solution of the correlation (fluctuating) Green function 
$\boldsymbol{v} \left( \tau, t \right)$ is governed 
by the non-equilibrium fluctuation-dissipation relation in the time domain 
\cite{PRL2012,NJP2010,ANNP2012}, 
\begin{align}
\boldsymbol{v}\left(\tau,t\right)  =\int_{t_{0}}^{\tau} \!\!\! d\tau_{1}\int_{t_{0}}^{t} \!\!\! d\tau_{2} 
\boldsymbol{u}\left( \tau,\tau_{1}\right)  \widetilde{\boldsymbol{g}}\left(  \tau_1,\tau_2\right) 
\boldsymbol{u}^{\dag}\left(  \tau_2,t_{0}\right)  \label{v(t)} ,
\end{align}
which is the general solution of Eq.~(\ref{vte}).  Similarly, for the open system with a single particle 
in the state with energy $\varepsilon_s$, we have $v(  t_s, t_s  )  = \int d\epsilon \chi(\epsilon)$  
in the steady-state limit $t_s  \rightarrow  \infty$, where \cite{Sci2015b,Sci2015a}
\begin{align}
\chi(\epsilon)=  [\mathcal{D}_b(\epsilon,t_s)+\mathcal{D}_d(\epsilon)]  f(\epsilon, {T} ) ,   \label{gfdt}
\end{align}
with $\mathcal{D}_b(\epsilon,t_s)=\sum_{j,k}\frac{J(\epsilon) \mathcal{Z}_{b_{j}} \mathcal{Z}_{b_{k}} }
{(\epsilon-\epsilon_{b_{j}})(\epsilon-\epsilon_{b_{k}})} {\cos [( \epsilon_{b_{j}} - \epsilon _{b _{k}} ) ( t _{s} - t _{0} ) ]}$
and $f(\epsilon, {T} )$ being the Bose-Einstein or Fermi-Dirac distribution function, depending on the system
being made of bosons or fermions. Equation (\ref{gfdt})  is the generalized fluctuation-dissipation theorem
modified by the localized bound states of open systems.  If there is no localized bound state, 
$\mathcal{Z}_{b_{j}}=0$, then $\chi(\epsilon)= \mathcal{D}_d(\epsilon)  f(\epsilon, {T} )$. 
This reproduces the standard equilibrium fluctuation-dissipation theorem at arbitrary temperature 
in the particle number representation. 

As a result, the steady-state average particle number in the system is given by 
 \begin{align}
 \langle a^{\dagger}(t_s) a(t_s)  \rangle
&  =|u\left(  t_s,t_{0}\right)|^2  n(t_0)   + v\left(t_s,t_s\right),  
\end{align}
where $n(t_0)$ is the initial particle number in the system.
If there is no localized bound state, $\mathcal{Z}_{b_{j}}=0$, the above result is simply reduced to
\begin{align} \langle a^{\dagger}(t_s) a(t_s)  \rangle=v ( t _{s} , t _{s} ) = \! \int \!\! 
d \omega \mathcal{D} ( \epsilon ) {f} ( \epsilon , T ) .
\end{align}
as a manifestation of the equilibrium fluctuation-dissipation theorem \cite{Caldeira1983}. 
In the Markovian limit  where $J(\epsilon)$ is a constant, which corresponds to the white band limit (or white noise), 
we further have \cite{Xiong2010}  
\begin{align}
\langle a^{\dagger}(t_s) a(t_s)  \rangle=v ( t _{s} , t _{s} ) \rightarrow f ( \epsilon_s, T ).  \label{mar_l}
\end{align}
This provides the foundation of statistical mechanics at equilibrium.

\begin{figure}
\center{\includegraphics[width=0.85\textwidth]{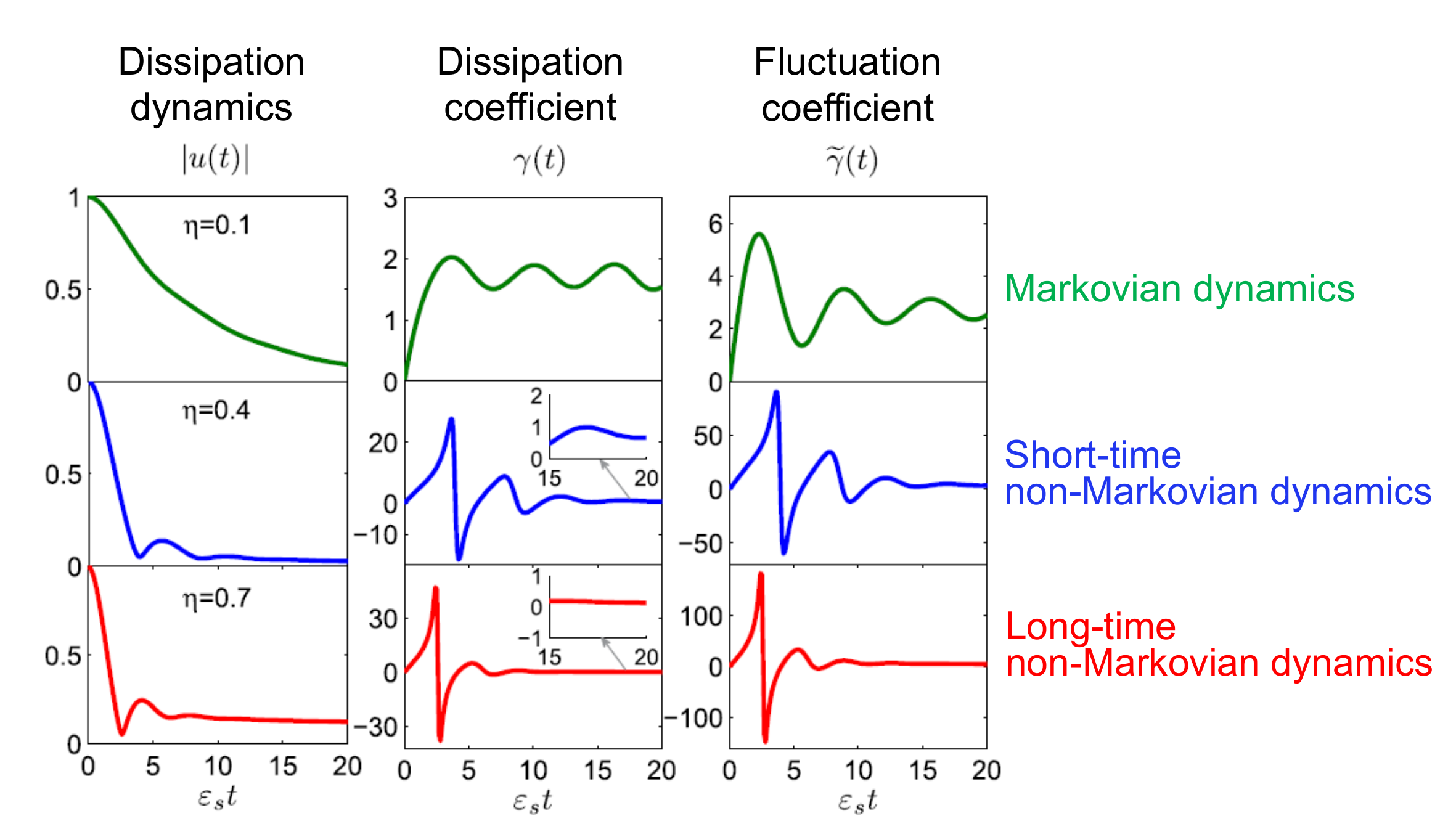}}
\caption{The non-Markovian dynamics in a bosonic system coupled to a sub-Ohmic reservoir
$J(\epsilon)= 2\pi\eta \epsilon (\epsilon/\epsilon_c)^{s-1}e^{-\epsilon/\epsilon_c}$ with $s=1/2$ \cite{PRL2012}. 
The energy cut-off $\epsilon_c=\varepsilon_s$ and the initial reservoir temperature 
$k_BT=\varepsilon_s$. Here we have taken $t_0=0$ so that we define $u(t,t_0)= u(t)$, $\gamma(t,t_0)=
\gamma(t)$ and $\widetilde{\gamma}(t,t_0)=\widetilde{\gamma}(t)$.}
\label{fig1}       
\end{figure}

With the analytical solutions (\ref{u(t)}) and (\ref{v(t)}), one can depict the dissipation  
and fluctuation dynamics through the time-dependent dissipation  and fluctuation coefficients, 
${\gamma}(t,t_0)$ and $\widetilde{\gamma}(t,t_0)$ in the master equation (\ref{Exact-ME}),
from which one can also find the solution of the reduced density matrix $\rho(t)$ if it is needed. 
Now we can give the general answer to the non-Markovian memory dynamics in open 
quantum systems \cite{PRL2012}: The nonexponential decays, the second terms in
Eq.~(\ref{u(t)}), is induced by the discontinuity in the 
imaginary part of the environmental-induced self-energy correction to the system, 
$\Sigma ( \epsilon \pm i 0 ^{+} ) = \Delta ( \epsilon ) \mp i \pi J ( \epsilon )$.
Depending on the detailed spectral density structure of $J(\omega)$, it could result in  
damping coefficients oscillating between positive and negative values in short times, as a short-time 
non-Markovian memory effect \cite{PRA2015}. The dissipationless oscillations, characterized 
by the localized bound states which are mainly arisen from band gaps or a finite band structure of 
environment spectral densities, provide a long-time non-Markovian memory effect. 
Fluctuation dynamics induces similar non-Markovian dynamics as dissipation through 
the generalized nonequilibrium fluctuation-dissipation relations that are obtained from 
nonequilibrium correlation Green function of Eq.~(\ref{v(t)}).  As an example, a bosonic 
open system in contact with a sub-Ohmic reservoir is shown in Fig.~\ref{fig1}, in which 
various non-Markovian dynamics discussed above show up.  Interesting exact numerical results
on spin-boson model with sub-Ohmic reservoir may be found from Ref.~\cite{Thorwart2010}. 

We have also introduced a quantitative measure of non-Markovian dynamics 
in terms of  two-time correlation functions in the same framework \cite{PRA2015}:
\begin{align}
\nonumber
{\cal N}(t,\tau) =& \Bigg| \frac{\langle A(t)B(t+\tau)\rangle}{\sqrt{\langle A(t)B(t)\rangle 
\langle A(t+\tau)B(t+\tau)\rangle}} \\
& -\frac{\langle A(t)B(t+\tau)\rangle_{\textrm BM}}{\sqrt{\langle A(t)B(t)\rangle_{\textrm BM} 
\langle A(t+\tau)B(t+\tau)\rangle_{\textrm BM}}}
\Bigg|,
\label{GennonM}
\end{align}
where $A$ and $B$ are two physical observables of the system. The exact two-time correlation function
$\langle A(t)B(t+\tau)\rangle$ can be obtained either from experiment or theoretical calculations, and
the two-time correlation function $\langle A(t)B(t+\tau)\rangle_{\textrm BM}$ can be evaluated through 
the BM master equation. For example, 
exact two-time current-current correlation $\langle I(t) I(t+\tau) \rangle$ in nano-electronic systems has been 
theoretically calculated \cite{Yang2014} and has recently been measured experimentally 
\cite{CurrentCurrent}. The two-time particle number correlation $\langle n(t)n(t+\tau) \rangle$ for
photonic systems is also experimentally measured through photon bunching and antibunching experiments 
\cite{antibunching1,antibunching2}, and the exact theoretical calculation is carried out in our recent work \cite{Ali2017}. 
More two-time intensity correlation functions have been experimentally 
measured in optical measurements \cite{malik,Livet,sutton}. While, the calculation of two-time correlation function 
$\langle A(t)B(t+\tau)\rangle_{\textrm BM}$ is rather simple under Born-Markovian approximation. Detailed 
discussions can be found from  \cite{PRA2015}. Thus, a quantitative measure of non-Markovian dynamics
can be given through two-time correlation functions which provide the direct physical picture of memory 
dynamics in open quantum systems.

\begin{figure}
\center{\includegraphics[width=0.8\textwidth]{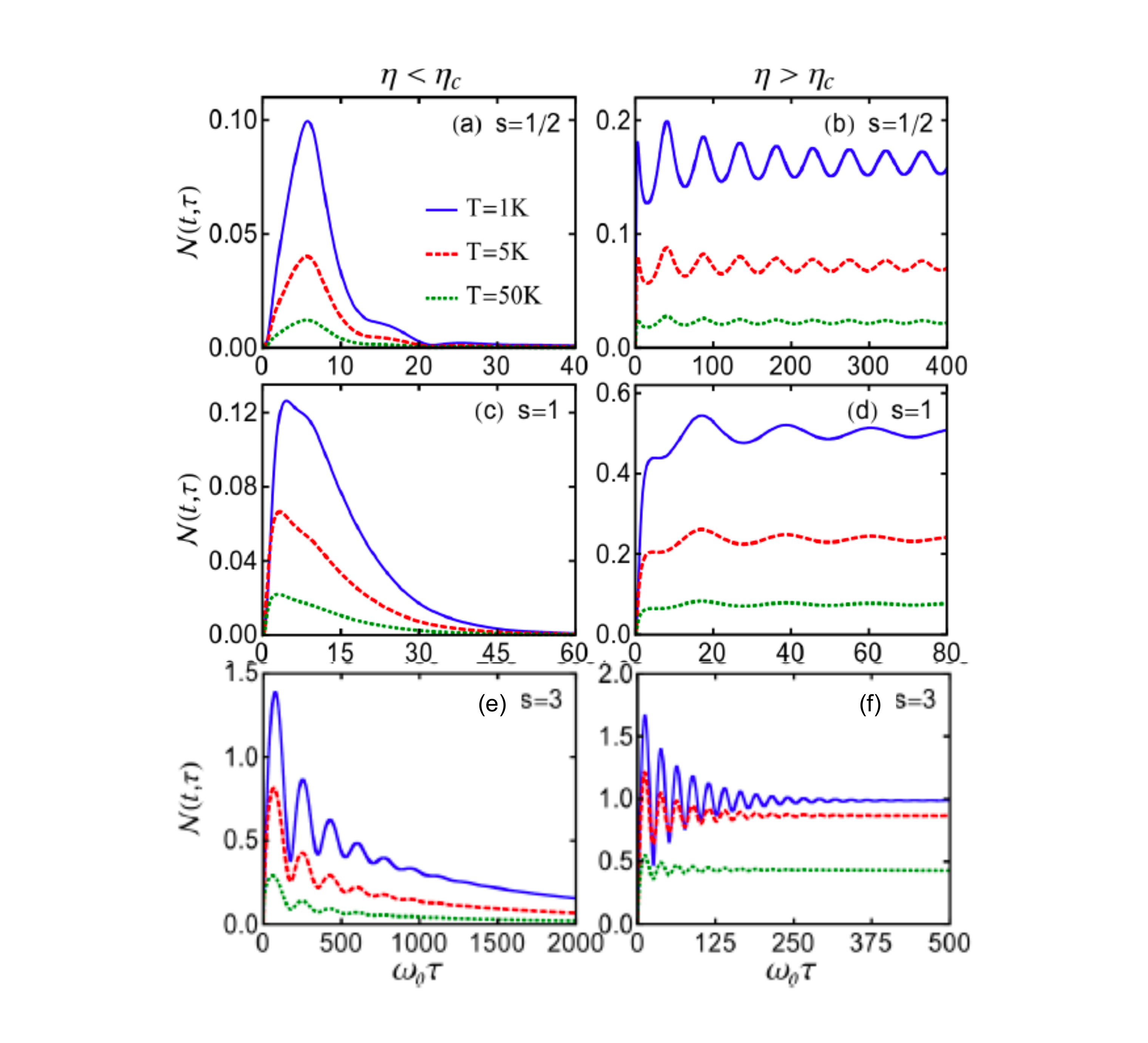}}
\caption{The measure of non-Markovian dynamics in terms of two-time correlation function \cite{PRA2015} 
for a bosonic system coupled to a thermal reservoir with Ohmic-type spectral density
$J(\epsilon)= 2\pi\eta \epsilon (\epsilon/\epsilon_c)^{s-1}e^{-\epsilon/\epsilon_c}$. 
The energy cut-off $\epsilon_c=5\varepsilon_s$ and the system is initially in Fock state with 
$n=1$, and $t=0$. As it is shown that when the system-reservoir coupling $\eta <\eta_c$,
no localized bound state exists. Correspondingly, the transient two-time correlation function 
shows quantitatively that memory effect vanishes when the system approaches to the thermal equilibrium state; 
while in the strong coupling regime $\eta > \eta_c$, the existence of the localized bound state 
prohibits the system to approach to the equilibrium state, and the non-Markovian dynamics (memory 
effect) can remains forever. }
\label{fig2}       
\end{figure}

As an illustration, in Fig.~\ref{fig2}, we present the non-Markovian dynamics
measure through the two-time correlations of the first-order photon coherence function, $\langle a^\dag(t)
a(t+\tau)\rangle$, as it precisely characterizes the long-time memory processes arisen from localized 
bound states ($\eta > \eta_c$) and the short-time memory processes resulted from nonexponential decays ($\eta <\eta_c$). 
The memoryless processes in the very weak system-environment coupling regime emerged 
naturally, and the exact master equation is then reduced to corresponding Born-Markov master equation \cite{Xiong2010}.
The above general non-Markovian dynamical properties of open quantum systems have also been applied
to investigate photonic dynamics in photonic crystals \cite{Lei2011,Sci2015b}, nonequilibrium 
photon statistics \cite{Ali2017}, nonequilibrium quantum phase transition \cite{Lin2016}, 
complexity of quantum-to-classical transition \cite{Sci2015a}; decoherence dynamics
of Majorana fermion in topological systems \cite{PRB18}, and quantum thermodynamics
in zero-dimensional systems very recently \cite{Ali2018}.  On the other hand, the exact master equation theory has
also been applied to study various transient quantum transport  physics in nanostructures
\cite{NJP2010,ANNP2012,PRB2015,Yang2014,Tu2012,Tu2014,Liu2016,Yang2017,Yang2018}

\section{Conclusions}
In conclusion, we have developed the exact master equation for a large class of non-interacting open 
quantum systems \cite{PRL2012}, including boson systems \cite{ANNP2012} and fermion systems 
\cite{PRB2008,NJP2010} as well as topological systems \cite{PRB18}.  The extension to spin systems 
is also in progress \cite{Yao2018}. We established the explicit connection of dissipation 
and fluctuation dynamics, described by the dissipation and fluctuation coefficients in the exact master 
equation, with nonequilibrium dynamics in terms of nonequilibrium Green functions. This connection
is crucially important because one can then apply the following conclusion to study non-Markovian 
dynamics of arbitrary interacting 
open systems, as long as the nonequilibrium Green functions are computable. The conclusion on 
general non-Markovian dynamics is summarized as follows \cite{PRL2012}: the general non-Markovian dynamics 
is embedded in the time convolution integro-differential equation of the nonequilbrium Green functions. 
In particular, the dissipation dynamics is fully determined by the propagating Green function. The general 
solution of the propagating Green function consists of nonexponential decays and dissipationless oscillations 
(localized bound states in open quantum systems) \cite{PRL2012}, which is indeed a universal
property of the propagating Green functions in arbitrary interacting many-body systems, according 
to the general principle of quantum field theory \cite{Peskin1995}. The non-exponential decays described by 
time-dependent decay rates oscillate between
positive (dissipation) and negative (back flowing) values in a short time, resulting in the short-time non-Markovian
dynamics \cite{PRA2015,Caldeira1983}. The localized bound states give dissipationless oscillations that make
the states of open systems depend forever to its initial state, as a long-time non-Markovian dynamics. 
Correspondingly, the open system is unable to approach to the thermal equilibrium state of the environment,
a property that initially noticed by Anderson long time ago \cite{Anderson1961} and is recently justified by
us in our exact master equation theory \cite{Sci2015a,Sci2015b}.  The fluctuating correlation (Keldysh) Green 
function, determined by the generalized nonequilibrium fluctuation-dissipation theorem [see Eq.~(\ref{v(t)})],
has the similar behaviors as the dissipation dynamics, namely it contains both the short-time and long-time
non-Markovian memory effects associated respectively with the continuous spectrum part and the localized bound state 
part of open systems.  This general picture of non-Markovian memory dynamics is fully
determined by the energy structures of the system and the environment, and the couplings between them, 
including also the initial states of the system and the environment, and is irrelevant to any mathematical
definition of non-Markovianity.

\section*{Acknowledgement}{}
I would like to thank my former and current students and Postdocs who have made the important contributions to the
development of this exact master equation theory for open quantum  systems, they include Matisse 
W. Y. Tu, J. Jin, C. U Lei, P. Y. Yang, P. Y. Lo, H. T. Tan, H. N. Xiong, H. L. Lam, and Y. W. Huang. I would also like to thank
my collaborators for the fruitful discussions in the related topics in the past years, they include Y. J. Yan, 
S. Gurvitz, A. Aharony, O. Entin-Wohlman, and F. Nori.  
This work is supported by the Ministry of Science and Technology of the Republic of China under the 
contract No. MOST 105-2112-M-006-008-MY3.

\end{document}